\documentclass[usenatbib,referee]{mn2e}
\usepackage{amsmath}
\usepackage{graphicx}
\usepackage{epsfig}
\usepackage{txfonts}
\usepackage{color}
\usepackage{natbib}
\usepackage{harpoon}

\title[Aspherical wind in symbiotic channel of SNe Ia]
{The effect of aspherical stellar wind of giant stars on the symbiotic channel of type Ia supernovae}
\author[C. Wu et al.]
{Chengyuan Wu$^{\rm 1}$\thanks{E-mail:wuchengyuan@mail.tsinghua.edu.cn}, Dongdong Liu$^{\rm 2,3,4}$\thanks{E-mail:liudongdong@ynao.ac.cn}, Xiaofeng Wang$^{\rm 1}$\thanks{E-mail:wang$\_$xf@mail.tsinghua.edu.cn}, Bo Wang$^{\rm 2,3,4}$\\
$^1$Physics Department and Tsinghua Center for Astrophysics (THCA), Tsinghua University, Beijing, 100084, China\\
$^2$Yunnan Observatories, Chinese Academy of Sciences, Kunming 650216, China\\
$^3$Key Laboratory for the Structure and Evolution of Celestial Objects, Yunnan Observatories, CAS, Kunming 650216, China\\
$^4$University of Chinese Academy of Sciences, Beijing 100049, China}

\begin{document}
\date{}
\pagerange{\pageref{firstpage}--\pageref{lastpage}} \pubyear{2019}
\maketitle

\label{firstpage}

\begin{abstract}\label{0. abstract}
The progenitor systems accounting for explosions of type Ia supernovae (SNe Ia) is still under debate. Symbiotic channel is one of the possible progenitor scenarios, in which the WDs in these systems increase in mass through wind accretion from their red giant companions. The mass-loss processes of the giants in the symbiotic systems could produce amount of circumstellar medium (CSM), and the detections of the interaction signals between SN ejecta and CSM can be used as an ideal way to distinguish the different progenitor models. However, the density distribution and geometric structure of the CSM around the symbiotic systems remain highly uncertain. By assuming that the tidal torque from binary interaction can increase the mass-loss rate of the red giant and cause the stellar wind concentrate towards the equatorial plane, we provide a simplified method to estimate the density and the degree of deviation from spherical symmetry of the CSM. Based on the calculations of the binary evolutions of symbiotic systems using stellar evolution code \texttt{MESA}, we obtained the parameter space for producing SNe Ia. We found that SNe Ia could originate from symbiotic systems with massive carbon-oxygen white dwarfs (CO WDs), while the half-opening angle of the stellar wind from red giant towards the WD varies with the binary evolution, resulting in the formation of surrounding CSM with different geometric structures. The corresponding properties of ejecta-CSM interactions may be examined by the spectropolarimetry observations in the future, from which one may find additional relationship between circumstellar environment of SNe Ia and their progenitor systems.
\end{abstract}


\section{Introduction} \label{1. Introduction}

Type Ia supernovae (SNe Ia) are successfully used as standardizable candles in measuring the cosmic distance (e.g. Riess et al. 1998; Perlmutter et al. 1999), and also have great contribution to the dynamical and chemical evolution of galaxies (e.g. Greggio \& Renzini 1983; Matteucci \& Greggio 1986). It has long been believed that SNe Ia are derived from thermonuclear explosions of carbon-oxygen white dwarfs (CO WDs; Hoyle \& Fowler 1960), however, the progenitor systems are still difficult to be identified (e.g. Wang \& Han 2012; Maoz, Mannucci \& Nelemans  2014; Livio \& Mazzali 2018; Wang 2018; Jha, Maguire \& Sullivan 2019; Ruiz-Lapuente 2019; Soker 2019a; Ruiter 2020). So far, at least two most popular progenitor models are widely discussed: (1) The single-degenerate (SD) model. In this model, a CO WD increases its mass by accreting material from its companion, and explodes as SN Ia when its mass approaches Chandrasekar-mass limit (${M}_{\rm Ch}$). Typically, the non-degenerate companion of the WD can be a main-sequence star (MS), sub-giant, red giant (RG), AGB star or a helium star (e.g. Whelan \& Iben 1973; Nomoto 1982; Li \& van den Heuvel 1997; Langer et al. 2000; Han \& Podsiadlowski 2004; Wang et al. 2009). (2) The double-degenerate (DD) model. In this model, supernova is originated from the coalescence of double CO WDs  (e.g. Iben \& Tutukov 1984; Webbink 1984), although the final fates of double WD merger may be somehow more complicated than what we thought (e.g. Nomoto \& Iben 1985; Schwab, Quataert \& Kasen 2016; Wu, Wang \& Liu 2019).

In the SD scenario, CO WD+RGB/AGB systems are thought to be progenitors of SNe Ia, in which CO WDs are able to increase in mass to reach ${M}_{\rm Ch}$ by accreting hydrogen-rich (H-rich) material from the stellar wind of their evolved giant companions (e.g. Boffin \& Jorissen 1988; Chen, Han \& Tout 2011; I{\l}kiewicz et al. 2019). This kind of wide binary systems, consisting of a cool, evolved giant and a hot, luminous WD or neutron star, are known as symbiotic stars (SySts). The evolved giants in symbiotic stars could be a normal red giant for S-type, or a Mira variable embedded in an optically thick dust shell for D-type (e.g. Milo{\l}ajewska 2012). Up to now, hundreds of symbiotic stars including several symbiotic recurrent nova (SyRN) systems such as RS Oph, T CrB and V407 Cyg are confirmed. The WDs in the SyRN systems are usually very massive or even close to ${M}_{\rm Ch}$, making the SyRNe to be promising progenitor candidates for SNe Ia.

Observationally, an important way to discriminate different progenitor models of SNe Ia is to detect the signal from the interaction between explosive ejecta of the supernova and the circumstellar medium (CSM). Generally, SNe that stem from the SD scenario should produce much denser CSM than the counterpart in the DD scenario. This is because in the SD scenario, binary mass-transfer process is indispensable before supernova explosion. The material ejected by the WD wind and the mass donor during the binary evolution may form the CSM of a certain density around the supernova. However, for the DD model, mass-transfer process stopped long before the occurrence of the supernova explosion due to the pretty long time-scale of the gravitational wave radiation, which means that there is relatively low CSM around the supernova (probably equals to the density of the interstellar medium, i.e. ${10}^{-24}\,\rm{g}\,\rm{cm}^{-3}$). Interaction of the SN ejecta with the CSM is able to produce electromagnetic radiation from X-ray to radio bands, depending on different radiation mechanism such as inverse Compton scattering and synchrotron emission (e.g. Chevalier 1982). Such signals have been detected in some cases. For instance, the first prototypical object for which circumstellar hydrogen has been detected was SN 2002ic, followed by SN 2005gj and SN 2008J (e.g. Hamuy et al. 2003; Aldering et al. 2006; Taddia et al. 2012). These SNe Ia show clear evidences for dense CSM, which indicates that their progenitors may be the SySts. SN 2006X was the first SN Ia which was found to show variable circumstellar Na I absorption lines in the spectra, and its progenitor is deduced to be a WD+RG system (e.g. Patat et al. 2007). The first case with both circumstellar interaction and time-varying narrow absorption lines was PTF 11kx, and in this system, multiple shells of CSM are detected, which can be described by a symbiotic nova progenitor, similar to RS Oph (e.g. Dilday et al. 2012). However, some other literatures claim that the CSM mass in PTF 11kx is larger than that predicted from symbiotic channel, which indicated that this SN may result from core-degenerate scenario (e.g. Soker et al. 2013).  More recently, SN 2015cp and SN 2018fhw are also reported to show H$\alpha$ emission lines likely resulting from ejecta-CSM interaction, which can be attributed to the SD progenitor origins (e.g. Graham et al. 2019; Vallely et al. 2019; Kollmeier et al. 2019). But it needs to be mentioned here that the inferred hydrogen mass in SN 2018fhw may significantly less than what theoretical predicted in SD scenario, which implies that alternative explanations for the origination of hydrogen in SN 2018fhw may needed such as the SN Ia ejecta unbinds a gas giant planet (e.g. Soker 2019b).

The CSM properties such as the density distribution and the value of half-opening angle ($\alpha$) can effect the observational features significantly (e.g. Moriya et al. 2013; Nagao, Maeda \& Ouchi 2020). Previous studies usually assumed a spherical stellar wind that is lost by the giant donor of the WD (e.g. Chomiuk et al. 2012; Meng \& Han 2016; Lundqvist et al. 2020), however, observations indicated that the giants in SySts could have aspherical stellar wind. For example, (1) the outburst of RS Oph in 2006 suggested an asymmetric shock wave which may result from the interaction between the nova ejecta and an equatorial enhancement in the red giant wind (e.g. O'Brien et al. 2006); (2) Spectropolarimetry observations of SN 2002ic suggested that its circumstellar environment may be dense and clumpy with disc-like geometry (e.g. Wang et al. 2004); (3) The multiple components of CSM presented in PTF 11kx implied a nonuniform material structure concentrated in the orbital plane (e.g. Dilday et al. 2012); (4) Many SNe Ia with high photospheric velocities (HVs) show significant excess flux in blue band at 40-100 days after their maximum light, which could be explained by the light echoes from the asymmetric shell or disc-like (model dependent) CS dust (e.g. Wang et al. 2019). Besides, mass-loss rates and wind velocities of the red giants remain highly uncertainties. Typically, the red giants in binary systems may have higher mass-loss rate than those in the field as the tidal interaction cannot be ignored. Comparing with field giants, Zamanov et al. (2008) found that the rotational velocities of giants in SySts are faster (i.e. by 1.5-4 times) and the mass-loss rates are higher (i.e. by 3-30 times), which could be attributed to the tidal interaction. In addition, metallicity can also alter the mass-loss rates of the giants in SySts and hence affect the birthrates of SNe Ia expected from the SySt channel. For example, (1) SNe Ia-CSM samples are almost from the late-type spiral galaxies, which implies that these objects may relate to young stellar population (e.g. Silverman et al. 2013). (2) The blue excess resulted from light echoes mentioned above is only observed in HV SNe Ia that may relate to metal-rich environment (e.g. Wang et al. 2013; Pan et al. 2015; Pan 2020; Wang et al. 2019).

The wind-accretion process and the expected parameter space of SNe Ia in SySts still remain uncertain. In present work, we plan to examine the effect of aspherical wind of giant that resulted from the tidal torque from the WD on the SySt channel leading to SNe Ia. According to our assumption, the WD in SySt can accrete material from wind more effectively due to higher density of the wind surrounding the WD. We calculated the parameter space for producing SNe Ia in both solar and metal-poor environments based on our revised wind prescription. Our article is organized as follows: In Sect. 2, we provide our basic assumptions and methods for the numerical simulations. The results of our simulations are shown in Sect. 3. Finally, we present the discussion in Sect. 4 and brief summary in Sect. 5.

\section{Numerical Methods}\label{Methods}

In order to obtain the evolutionary properties of the binaries in the assumptions of aspherical wind and the predicted parameter space for producing SNe Ia, we employ the stellar evolution code \texttt{MESA} (version: 10398; Paxton et al. 2011, 2013, 2015, 2018) to calculate the evolutions of WD+FGB/AGB binaries with solar metallicity (with Z=0.02) and in the metal-poor environment (with Z = $1\times10^{-4}$). The CO WDs are treated as point masses, with which masses of WDs (${M}_{\rm WD}$) are between $1.0{M}_\odot$ and $1.1{M}_\odot$ (the very massive CO WD evolved from single star evolution). During the evolution, WDs can accrete portion of wind material, leading to the mass growth. We assume that WDs explode as SNe Ia when their mass grow to $1.378{M}_{\odot}$. In our simulations, the masses of MS companions ${M}_{\rm com}$ vary from 1.5 to 4.0 ${M}_\odot$ with mass intervals of $\Delta\,{M}_{\rm com}=0.1$, and the initial orbital periods (${P}^{\rm i}$) are in the range of ${\rm log}({P}^{\rm i}/{\rm day})=1.0-4.0$ with period intervals of $\Delta\,{\rm log}({P}/{\rm day})=0.1$.

\subsection{Mass loss of the giants}

Owing to the tidal force exerted on the companion stars from the WDs, stellar wind from the giant may deviate from spherical symmetry, leading to the formation of aspherical CSM around the binary systems. Zahn (1975) investigated the dynamical tide in close binaries, and found that the torque in tidal friction is related to $(R/{R}_{\rm L})^{6}$, where ${R}_{\rm L}$ is the Roche Lobe radius of the mass donor.

Based on the tidal interaction in the close binaries, Tout \& Eggleton (1988) proposed tidal enhanced wind model, in which the mass-loss rate of the giant can be expressed as
\begin{equation}
\dot{M}=\dot{M}_{\rm wind}\cdot\{1+B\times\,{\rm min}[(\frac{R}{{R}_{\rm L}})^{6},\frac{1}{{2}^{6}}]\},
\end{equation}
where $\dot{M}_{\rm wind}$ is the wind mass-loss rate of the RG in single star system and B is a free parameter to control the effect of the tidal enhancement. However, they did not consider the effect of wind material concentrating towards the orbital plane. Here, we assumed that the tidal force could cause (1) the increase in mass-loss rate of the donor, (2) the stellar wind of the donor to be concentrated towards the equatorial disc. The assumption of enhancement in mass-loss rate of the donor is same as Tout \& Eggleton (1988). The deflection angle ($\theta$) is defined as ($\frac{\pi}{2}-{\alpha}$), where ${\alpha}$ is the half-opening angle of the conical stream from the giant to the WD, and we assumed that ${\rm sin}\theta$ may obey three different function forms as follows in our calculations.

(1) linear function:
\begin{equation}
{\rm sin}\theta={a}_{1}\times(\frac{R}{{R}_{\rm L}})^{6}+{b}_{1};
\end{equation}

(2) binomial function:
\begin{equation}
{\rm sin}\theta={a}_{2}\times(\frac{R}{{R}_{\rm L}})^{12}+{b}_{2}\times(\frac{R}{{R}_{\rm L}})^{6};
\end{equation}

(3) exponential function:
\begin{equation}
{\rm sin}\theta={a}_{3}\times{\rm exp}\,[{b}_{3}\times(\frac{R}{{R}_{\rm L}})^{6}].
\end{equation}
The free parameters of tidal effect can be constrained from the observations. The RS Oph is a symbiotic system which involves a $1.33{M}_\odot$ WD, and a $0.8{M}_\odot$ RG companion. The orbital period and separation of this system are around 453.6$\,$days and 1.48AU, respectively (e.g. Brandi et al. 2009; Booth, Mohamed \& Podsiadlowski 2016). Mohamed, Booth \& Podsiadlowski (2010) did 3D hydrodynamic simulations of the RS Oph system, by assuming that the radius of RG is approximated to 0.8 times of its the Roche-lobe radius, and found that the half-opening angle of the wind material to the WD is about $30^{\circ}$. By adopting their results, we assumed that ${\rm sin}\theta=\frac{\sqrt{3}}{2}$ when the radius of a RG expands to $0.8{R}_{\rm L}$. Besides, ${\rm sin}\theta$ should comply with additional boundary conditions, i.e. (1) when $R<<{R}_{\rm L}$, ${\rm sin}\theta$=0; (2) when ${R}\sim{R}_{\rm L}$, ${\rm sin}\theta\approx1$. Substituting the boundary conditions into the three different function forms of ${\rm sin}\theta$, we could constrain the corresponding parameters, i.e.

for the linear function:
\begin{equation}
{\rm sin}\theta=3.3\times(\frac{R}{{R}_{\rm L}})^{6};
\end{equation}

for the binomial function:
\begin{equation}
{\rm sin}\theta=-3.137\times(\frac{R}{{R}_{\rm L}})^{12}+4.137\times(\frac{R}{{R}_{\rm L}})^{6};
\end{equation}

for the exponential function:
\begin{equation}
{\rm sin}\theta=39.3\times{\rm exp}\,[-(\frac{R}{{R}_{\rm L}})^{-6}].
\end{equation}

\begin{figure}
\begin{center}
\includegraphics[width=0.65\textwidth]{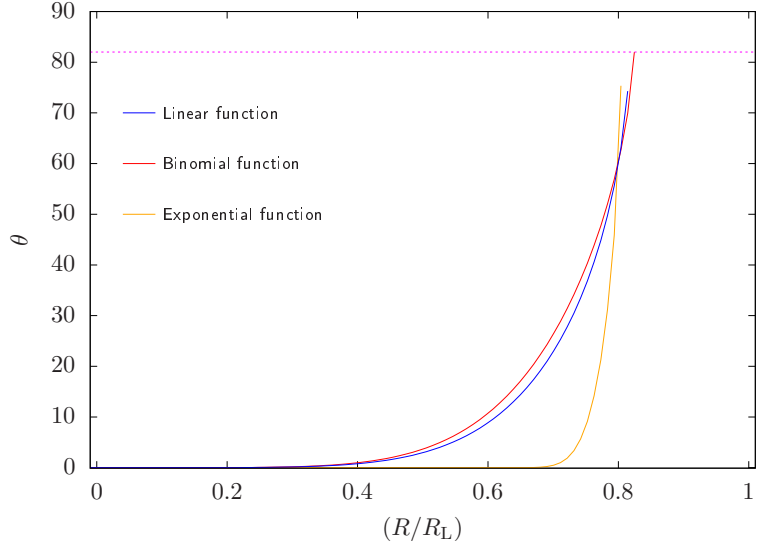}
 \caption{The varition of wind deflection angle ($\theta$) with ${R}/{R}_{\rm L}$. Solid carves represent three different functional relationships between ($\theta$) and $(\frac{R}{{R}_{\rm L}})^{6}$, whereas dotted line represents the upper limit of deflection angle.}
  \end{center}
\end{figure}

Fig.\,1 shows the variation of ${\rm sin}\theta$ with $\frac{R}{{R}_{\rm L}}$ in three different functional forms. Note that the maximum value of ${\rm sin}\theta$ cannot greater than 1, hence, we artificially limit the maximum value of ${\rm sin}\theta=0.99$, and assumed that the stellar wind is concentrated in the equatorial disc if ${\rm sin}\theta>0.99$. From the figure, we can see that when ${R}<0.8{R}_{\rm L}$, $\theta$ in linear and binomial forms have the similar increase rate, whereas $\theta$ changes rapidly in the exponential form when ${R}>0.7{R}_{\rm L}$. In the current work, we assumed that ${\rm sin}\theta$ is satisfied with binomial form, and discuss the uncertainties in Sect.\,4.

\subsection{Mass accretion of WDs}

When the companion enters into the giant branch, portion of the stellar wind material could be accreted by the WD. Here, we assumed three different accretion modes of the WD as follows.

(1) when the mass donor is in RGB phase (mass fraction of hydrogen in the centre ${H}_{\rm centre}$ is less than ${10}^{-6}$, and carbon in the centre (${C}_{\rm centre}$) is greater than ${10}^{-6}$), the accretion rate of WD is in the Bondi-Hoyle accretion form (e.g. Bondi \& Hoyle 1944; Boffin \& Jorissen 1988) with an extra enhancement in the density of the wind material, which is
\begin{equation}
\dot{M}_{\rm acc, BH}=-\frac{1}{\sqrt{1-{e}^{2}}}\,(\frac{G{M}_{\rm WD}}{{v}_{\rm w}^{2}})^{2}\,\frac{{\alpha}_{\rm acc}\dot{M}_{\rm loss}}{2{a}^{2}(1+\frac{{v}_{\rm orb}^{2}}{{v}_{\rm w}^{2}})^{1.5}}\cdot\frac{\pi}{2\alpha},
\end{equation}
where ${\alpha}_{\rm acc}$ is accretion efficiency parameter which is set to be $1.5$ (a default value in MESA), ${v}_{\rm orb}=\sqrt{\frac{G({M}_{\rm com}+{M}_{\rm WD})}{a}}$ is the orbital velocity, ${v}_{\rm w}=\frac{1}{2}\sqrt{\frac{G{M}_{\rm com}}{{R}_{\rm com}}}$ is the wind velocity, and $\alpha$ is the half-opening angle of the conical stream from the giant to the WD.

(2) Recent hydrodynamical simulations suggested that if the wind acceleration radius of AGB star is larger than the Roche-lobe radius of itself in binary system, the material in the wind acceleration zone could be accreted by the WD through the inner Lagrangian point, which cause the formation of wind Roche-lobe overflow mechanism (WRLOF; see Mohamed \& Podsiadlowski 2007). WRLOF may occur if the velocity of AGB wind is close to the orbital velocity of the WD. Here, we assume that when the mass donor evolves to AGB phase (Both central carbon and hydrogen are less than ${10}^{-6}$), and if $\frac{1}{3}<\frac{{v}_{\rm orb}}{{v}_{\rm w}}<3$, the WD accretes material from the wind of the donor in WRLOF mode with an extra enhancement in the density of the wind material, in which the accretion rate can be expressed as
\begin{equation}
\dot{M}_{\rm acc, WRLOF}=\frac{25}{9}{q}^{2}[-0.284(\frac{{R}_{\rm d}}{{R}_{\rm L}})^{2}+0.918(\frac{{R}_{\rm d}}{{R}_{\rm L}})-0.234]\cdot\frac{\pi}{2\alpha}\cdot\dot{M}_{\rm loss},
\end{equation}
where, $q=\frac{{M}_{\rm WD}}{{M}_{\rm giant}}$ is the mass ratio of WD to companion, ${R}_{\rm d}=\frac{{R}_{\rm giant}}{2}\times(\frac{{T}_{\rm d}}{{T}_{\rm eff, giant}})^{-2.5}$ is the dust radius of the AGB star, and ${T}_{\rm eff, giant}$ is the temperature of the giant, ${T}_{\rm d}=1500K$ is the temperature of the carbon-rich dust (see Abate et al. 2013).

(3) When ${\rm sin}\theta\geq0.99$, we assumed that the stellar wind is in an equatorial disc form, in which the accretion rate of WD is related to the crossing area of its gravitational radius in the orbital plane. The accretion rate is approximately equals to
\begin{equation}
\dot{M}_{\rm acc, DISC}=\frac{{R}_{\rm G}{v}_{\rm orb}}{{\pi}a{v}_{\rm w}}\dot{M}_{\rm loss},
\end{equation}
where ${R}_{\rm G}=\frac{2G{M}_{\rm WD}}{{v}_{\rm w}^{2}+{v}_{\rm orb}^{2}}$ is the gravitational radius of the WD (e.g. Lv et al. 2009).

Note that the WD cannot accrete all of the wind matter ejected by the companion, therefore, we artificially enforce the mass-accretion rate of the WD from those three accretion forms mentioned above as $\dot{M}_{\rm acc, BH}\leq0.5$, $\dot{M}_{\rm acc, WRLOF}\leq0.8$ and $\dot{M}_{\rm acc, DISC}\leq0.9$ (e.g. I{\l}kiewicz et al. 2019; Lv et al. 2009). In addition, if the red giant fills its Roche-lobe during the evolution, dynamical unstable mass transfer may occur since the mass-transfer rate increases dramatically, leading to the formation of common envelope. In order to avoid the complicate calculations, we artificially stop the simulations if mass-transfer rates from Roche-lobe overflow are greater than ${10}^{-3}\,{M}_{\odot}\,\rm{yr}^{-1}$, and assume that the corresponding systems are undergoing dynamically unstable mass transfer.

\subsection{Mass-loss rate estimation}

We assumed that the wind mass-loss rate of the giant relates to its mass, radius and luminosity. When the companion evolves to RGB phase, we adopt wind-loss rate formula from Reimers (1975), which is
\begin{equation}
\dot{M}_{\rm wind}=4\times{10}^{-13}{\eta}_{\rm R}\frac{LR}{M},
\end{equation}
where, ${\eta}_{\rm R}=0.5$ is a typical value in MESA. Here, the wind mass-loss rate is enhanced by tidal force, thus, the real mass-loss rate in RGB phase is
\begin{equation}
\dot{M}=-4\times{10}^{-13}\,{\eta}_{\rm R}\,\frac{RL}{M}\,\{1+B\times\,{\rm min}[(\frac{R}{{R}_{\rm L}})^{6},\frac{1}{{2}^{6}}]\},
\end{equation}
where R, L and M are radius, luminous and mass of the donor in solar units, respectively. For the SySt RS Oph, the radius of the RG is about $\sim59.1{R}_{\odot}$ which is obtained from the estimation of the average radius for the corresponding spectral type of the RG (e.g. Zamanov et al. 2007). The interval of recurrent nova eruption of RS Oph is about $20\,\rm{yrs}$, corresponding to a mass-accretion rate of about $\dot{M}_{\rm acc}=5\times{10}^{-8}\,{M}_\odot\,\mbox{yr}^{-1}$ (see Fig.\,4 of Hillman et al. 2016). To estimate the tidally enhanced parameter, we evolved binaries with systematic parameters similar to RS Oph. The mass-loss rate of a $0.8{M}_\odot$ star is about $\sim7.6\times{10}^{-9}\,{M}_{\odot}\,\rm{yr}^{-1}$ when its radius expands to the desired value without considering the effect of tidal enhancement, and the Roche-lobe radius of the RG companion in RS Oph system is approximated as $107{R}_\odot$ based on the estimation of Roche-lobe effective radius from Eggleton (1983). Thus, the wind-accreting efficiency of the WD is about $\sim30\%$ depending on the assumptions from Boffin \& Jorissen (1988). Therefore, we can estimate the tidally enhanced parameter ${B}$ is about a few thousand, and we adopted ${B}=4000$ in our simulations. We adopted wind mass-loss rate formula from Bl\"ocker (1995) when the companion evolves to AGB phase. The mass-loss rate can be expressed as
\begin{equation}
\dot{M}_{\rm Blocker}=4.83\times{10}^{-9}{\eta}_{\rm B}{M}^{-2.1}{L}^{2.7}\times{\dot{M}_{\rm Reimers}},
\end{equation}
where, the wind parameter ${\eta}_{\rm B}$ is set to be 0.1, which is a typical value recommended in MESA.

\subsection{Mass growth of WDs}

The WDs in our simulations are treated as point masses. In order to limit the mass growth of WDs, we adopted the prescription provided by Hachisu et al. (1999). In their prescriptions, the critical mass accretion rate of WD relates to the mass fraction of hydrogen (${\chi}_{\rm H}$) and WD mass (${M}_{\rm WD}$), i.e.
\begin{equation}
\dot{M}_{\rm cr, H}=5.3\times{10}^{-7}\frac{(1.7-{\chi}_{\rm H})}{{\chi}_{\rm H}}({M}_{\rm WD}/{M}_\odot-0.4){M}_\odot\,\mbox{yr}^{-1}.
\end{equation}
If the mass accretion rate is higher than $\dot{M}_{\rm cr, H}$, the WD accumulates hydrogen as a rate of $\dot{M}_{\rm cr, H}$, the unburned material is blown away by the optically thick wind (e.g. Kato \& Hachisu 1994; Hachisu, Kato \& Nomoto 1996). If $\dot{M}_{\rm acc}$ is lower than $\dot{M}_{\rm cr, H}$ but higher than $\frac{1}{8}\dot{M}_{\rm cr, H}$, the WD can accumulate all the accreted material. If $\dot{M}_{\rm acc}$ is lower than $\frac{1}{8}\dot{M}_{\rm cr, H}$, strong H-shell flash is able to occur on the surface of WD, and in this case, the WD cannot accumulate any material. According to the assumptions above, the mass-accumulation efficiency of hydrogen ${\eta}_{\rm H}$ can be expressed as follows:
\begin{equation}
{\eta}_{\rm H}=\left\{
\begin{array}{rcl}
\frac{\dot{M}_{\rm cr, H}}{\dot{M}_{\rm acc}}, & & {\dot{M}_{\rm acc}>\dot{M}_{\rm cr, H}},\\
1, & & {\frac{1}{8}\dot{M}_{\rm cr, H}\leq{\dot{M}_{\rm acc}}\leq{\dot{M}_{\rm cr, H}}},\\
0, & & {{\dot{M}_{\rm acc}}\leq{\frac{1}{8}\dot{M}_{\rm cr, H}}}.
\end{array} \right.
\end{equation}
Hence, the mass-growth rate of helium layer can be expressed as
\begin{equation}
\dot{M}_{\rm He}={\eta}_{\rm H}\dot{M}_{\rm acc}.
\end{equation}

Similarly, the mass-accumulation efficiency of He layer can be expressed as
\begin{equation}
{\eta}_{\rm He}=\left\{
\begin{array}{rcl}
\frac{\dot{M}_{\rm cr, He}}{\dot{M}_{\rm He}}, & & {\dot{M}_{\rm He}>\dot{M}_{\rm cr, He}},\\
1, & & {\dot{M}_{\rm st, He}\leq\dot{M}_{\rm He}\leq\dot{M}_{\rm cr, He}},\\
{\eta}_{\rm He}^{'}, & & {\dot{M}_{\rm low, He}<\dot{M}_{\rm He}<{\dot{M}_{\rm st, He}}},\\
0, & & {{\dot{M}_{\rm He}}\leq{\dot{M}_{\rm low, He}}}.
\end{array} \right.
\end{equation}
where, $\dot{M}_{\rm low, He}$ and $\dot{M}_{\rm cr, He}$ are the lowest and highest critical transfer rate, $\dot{M}_{\rm st, He}$ is the minimum accretion rate of stable He-shell burning, ${\eta}_{\rm He}^{'}$ is the mass-accumulation efficiency of helium in weak He-shell flashes. If $\dot{M}_{\rm He}$ is lower than $\dot{M}_{\rm low, He}$, a helium detonation may cause a supernova explosion, which is out of our consideration, and we assumed the mass-accumulation rate of He layer to be zero. If $\dot{M}_{\rm He}$ is higher than $\dot{M}_{\rm cr, He}$, the WD accumulates helium layer as a rate of $\dot{M}_{\rm cr, He}$, the rest of material is blown away by the optically thick wind. If $\dot{M}_{\rm He}$ is in the range of $\dot{M}_{\rm st, He}$ to $\dot{M}_{\rm cr, He}$, the He-shell burning is stable. If $\dot{M}_{\rm He}$ is in the range of $\dot{M}_{\rm low, He}$ to $\dot{M}_{\rm st, He}$, the mass-increase rate of CO core can be expressed as:
\begin{equation}
\dot{M}_{\rm CO}={\eta}_{\rm He}\dot{M}_{\rm He}={\eta}_{\rm H}{\eta}_{\rm He}\dot{M}_{\rm acc}.
\end{equation}
The critical transfer rates are
\begin{equation}
\dot{M}_{\rm low, He}=4.0\times{10}^{-8}\,{M}_{\odot}\,\mbox{yr}^{-1}
\end{equation}
and
\begin{equation}
\dot{M}_{\rm cr, He}=7.2\times{10}^{-6}\,({M}_{\rm WD}/{M}_{\odot}-0.6)\,{M}_{\odot}\,\mbox{yr}^{-1},
\end{equation}
respectively, based on the previous simulations of mass-accretion WD evolutions (e.g. Woosley, Taam \& Weaver 1986; Nomoto 1982). ${\eta}_{\rm He}^{'}$ relates to the WD mass and $\dot{M}_{\rm He}$, and we adopted the linearly interpolated from a grid computed by Kato \& Hachisu (2004).

For the low metallicity (Z=0.0001), we adopted the results of Chen et al. (2019), in which they provided comprehensive models of nova at metal-rich and metal-poor environment, for the upper and lower limits of steady hydrogen burning, i.e.
\begin{equation}
{\rm log}(\dot{M}_{\rm lower})=-12.21+13.32\times{M}_{\rm WD}-11.90\times{M}_{\rm WD}^{2}+3.82\times{M}_{\rm WD}^{3},
\end{equation}
\begin{equation}
{\rm log}(\dot{M}_{\rm upper})=-9.81+7.25\times{M}_{\rm WD}-5.17\times{M}_{\rm WD}^{2}+1.32\times{M}_{\rm WD}^{3}.
\end{equation}
In the metal-poor environment, we have not changed the polynomial fitting of ${\eta}_{\rm He}$ since the mass-accumulation rate of helium may be slightly influenced by the metallicity (e.g. Wu et al. 2017).

\subsection{Orbital angular-momentum losses}

During the binary evolution, mass-loss from both donor and WD are needed to be considered. The wind material which is lost from the surface of binary stars can take away the specific angular momentum of mass donor and CO WD. The angular-momentum-loss due to mass-loss from the donor is
\begin{equation}
\dot{J}_{\rm ML}({\rm donor})=-\dot{M}_{\rm d}(\frac{{a}{M}_{\rm WD}}{{M}_{\rm WD}+{M}_{\rm d}})^{2}\frac{2\pi}{{P}_{\rm orb}},
\end{equation}
and
\begin{equation}
\dot{J}_{\rm ML}({\rm WD})=-(1-\eta)\dot{M}_{\rm WD}(\frac{{a}{M}_{\rm d}}{{M}_{\rm WD}+{M}_{\rm d}})^{2}\frac{2\pi}{{P}_{\rm orb}}
\end{equation}
for the angular-momentum-loss from the vicinity of the CO WD, in which $\eta$ is the mass-accumulation rate of the CO WD.

\section{Numerical Results}\label{Results}
\subsection{Binary evolution calculations}

\begin{figure}
\begin{center}
\includegraphics[width=0.75\textwidth]{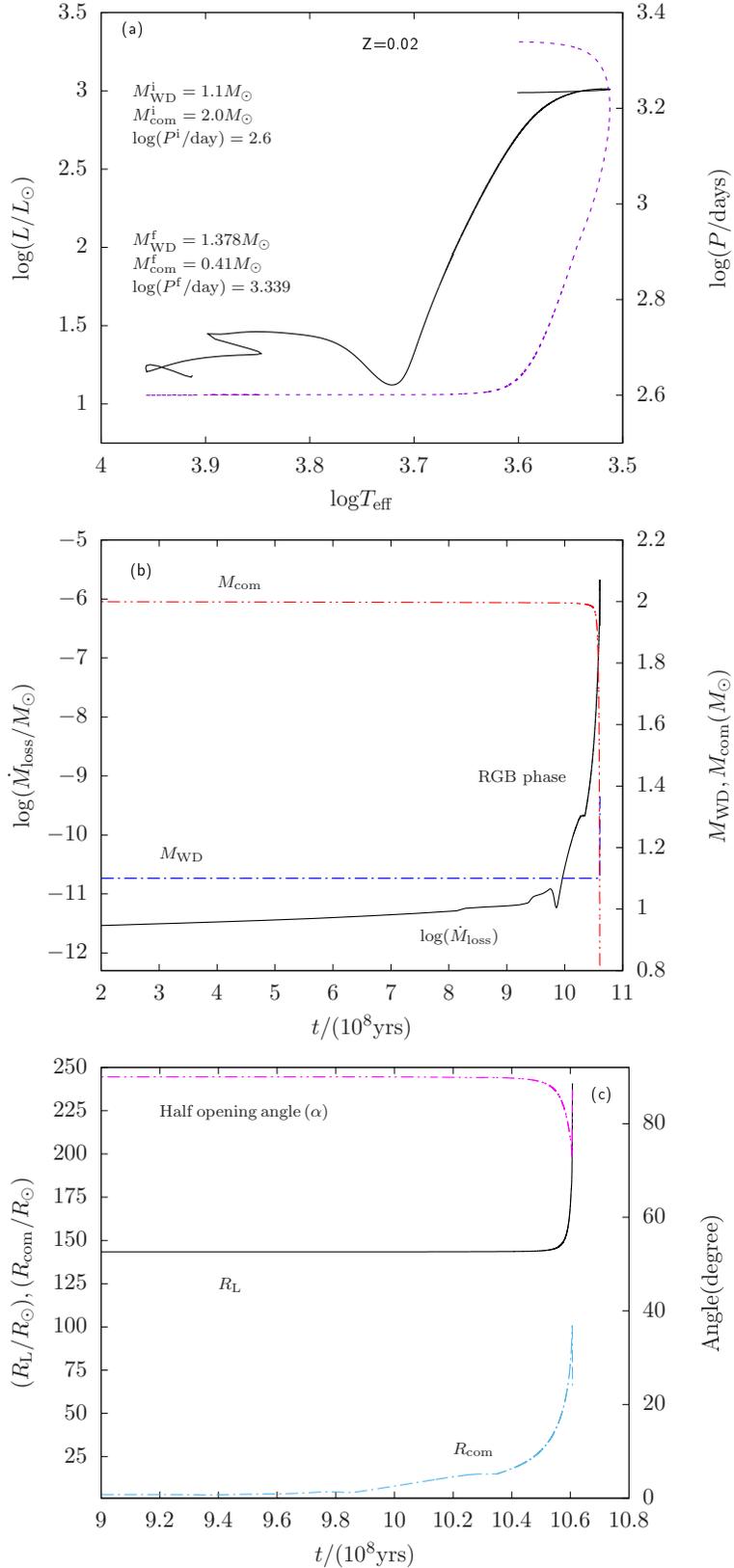}
 \caption{A typical binary evolution example for producing SN Ia through Bondi-Hoyle accretion mode in symbiotic channel with solar metallicity, in which ${M}_{\rm WD}^{\rm i}=1.1{M}_{\odot}$, ${M}_{\rm com}^{\rm i}=2.0{M}_{\odot}$ and ${\rm log}{P}^{\rm i}=2.6$, respectively. Panel (a): solid black curve represents the evolutionary track of the companion in the HR diagram, whereas the dashed line represents the evolutionary track of the orbital period of the binary system. Panel (b): solid black curve represents the evolution of the mass-loss rate of the companion star in logarithmic form from the end of main-sequence stage to the SN explosion. Red and blue dash-dotted curve show the evolution tracks of the masses of WD and companion; panel (c): solid black curve and blue dash-dotted curve show the evolution of Roche-Lobe, radius and the stellar radius of the companion, whereas the magenta dash-dotted curve represents the change of half-opening angle of the stellar wind material from the companion to the WD.}
  \end{center}
\end{figure}

Figure 2-5 present typical examples of binary evolutions, in which the masses of WDs could reach ${M}_{\rm Ch}$ ($1.378{M}_{\odot}$) through wind accretion. In Fig.\,2, the initial WD mass and orbital period are set to be ${M}_{\rm WD}=1.1{M}_{\odot}$ and ${\rm log}({P}^{\rm i}/{\rm day})=2.6$, respectively. The companion star is a $2.0{M}_{\odot}$ MS with solar metallicity (Z=0.02). The wind mass-loss rate is quite small (on the order of magnitude of around ${10}^{-11}\,{M}_{\odot}\,\rm{yr}^{-1}$) during the main-sequence and Hertzsprung gap phases. After about $1\times{10}^{9}$ years, the companion enters into RGB phase and begins to expand gradually, resulting in the increase of luminosity and mass-loss rate. According to the prescription described in Sect\,2, the WD could accretes portion of material in the stellar wind. However, the WD cannot accumulate mass until the mass-accretion rate increases to be greater than $1.0\times{10}^{-7}\,{M}_{\odot}\,\rm{yr}^{-1}$, since all the accreted material is ejected by the strong hydrogen flash on the surface of WD. Thus, the WD only increases in mass during the evolution of the last $1.8\times{10}^{6}\,\rm{yr}$, which can be seen in panel (b) of Fig.\,2. The mass-loss from the vicinity of the WD carries away orbital angular momentum, and the non-conservative mass-transfer process leads to the increase of orbital period. As shown in panel (a) of Fig.\,2, the binary system evolves to the long orbital period during the RGB phase, and finally reached ${\rm log}({P}^{\rm f}/{\rm day})=3.34$ when SN Ia explosion occurs. The evolution of radius of the companion is shown in panel (c) of Fig.\,2. As can be seen from this figure that the radius of companion and its Roche-lobe are $65.4{R}_{\odot}$ and $417.7{R}_{\odot}$, respectively. In this case, the half-opening angle of the stellar wind from the companion to the WD is about ${87.2}^{\circ}$, which means that the WD is accreting material in the spherical wind through Bondi-Hoyle accretion mode.

\begin{figure}
\begin{center}
\includegraphics[width=0.75\textwidth]{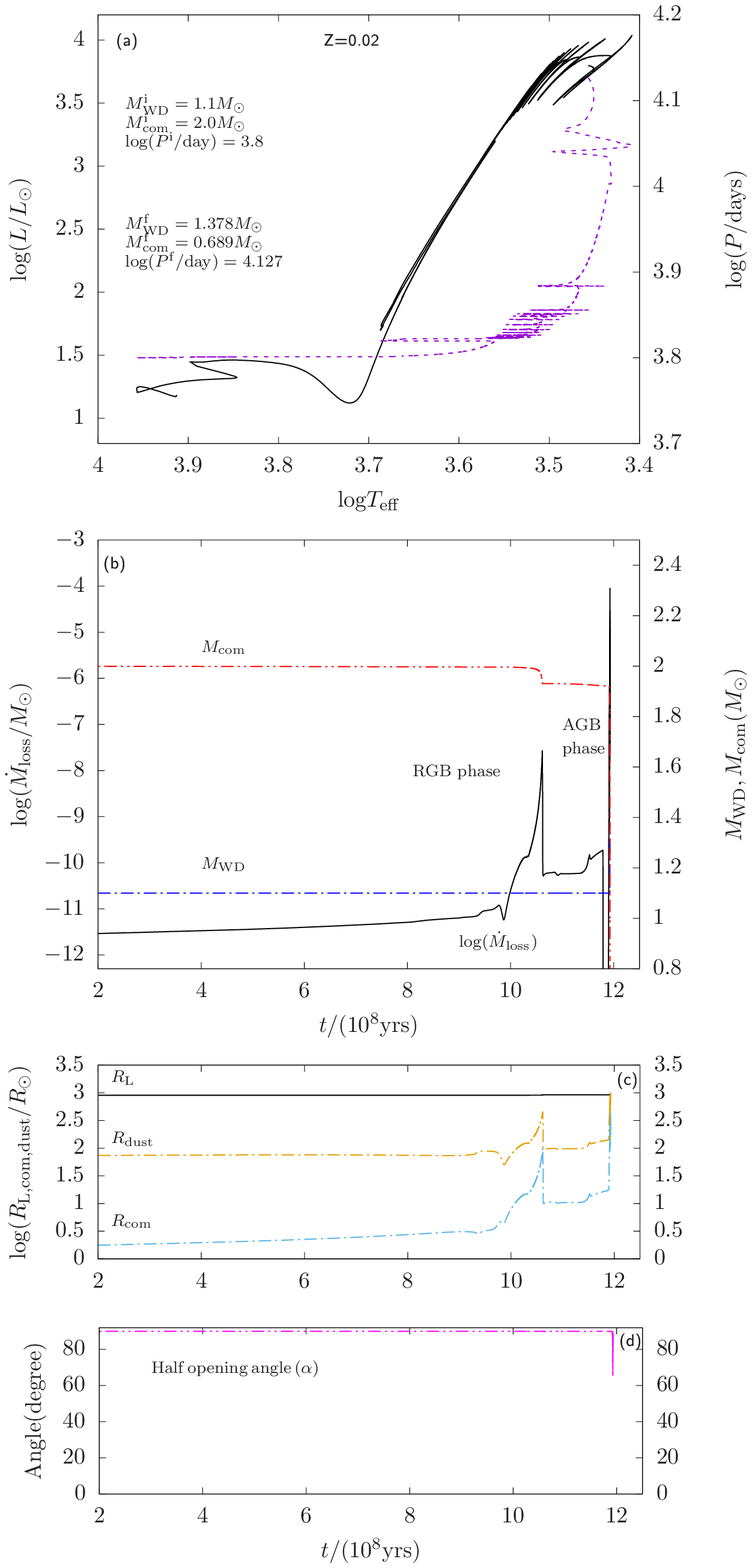}
 \caption{Binary evolution example for producing SN Ia through WRLOF accretion mode in symbiotic channel with solar metallicity, in which the binary comprises a $1.1{M}_{\odot}$ WD and a $2.0{M}_{\odot}$ MS with orbital period of ${\rm log}{P}^{\rm i}=3.8$.}
  \end{center}
\end{figure}

Fig.\,3 shows an example of SN Ia arising from AGB channel, where the initial companion mass ${M}_{\rm WD}=2.0{M}_{\odot}$ and initial orbital period is ${\rm log}({P}^{\rm i}/{\rm day})=3.8$, respectively. Due to the initially wide separation, the wind mass-loss rate of the companion is lower than ${10}^{-7}\,{M}_{\odot}\,\rm{yr}^{-1}$ in the whole RGB phase since the effect of tidal enhancement is relatively weak. However, when the companion evolves into AGB phase, the radius and mass-loss rate increase rapidly since the star has more effective mass-loss mechanism. In this case, the WD increases its mass to ${M}_{\rm Ch}$ in $4.5\times{10}^{5}\,\rm{yr}$ through WRLOF wind accretion mode.

\begin{figure}
\begin{center}
\includegraphics[width=0.75\textwidth]{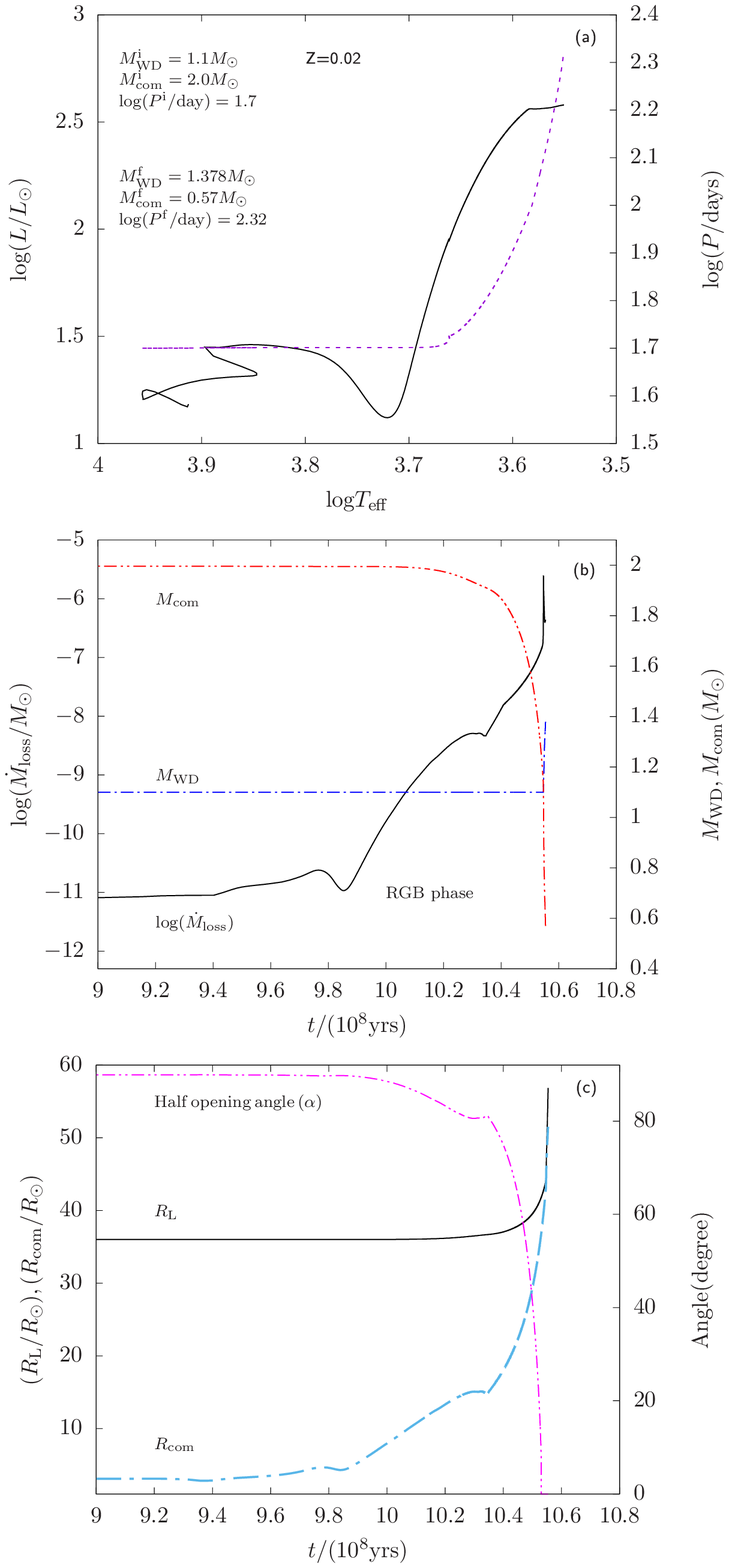}
 \caption{Binary evolution example for producing SN Ia through equatorial disc accretion mode in symbiotic channel with solar metallicity, in which the binary comprises a $1.1{M}_{\odot}$ WD and a $2.0{M}_{\odot}$ MS with orbital period of ${\rm log}{P}^{\rm i}=1.7$.}
  \end{center}
\end{figure}

In Fig.\,4, we present an example for producing SN Ia in which the wind material is concentrated in the equatorial disc. The initial companion mass and initial orbital period are ${M}_{\rm WD}=2.0{M}_{\odot}$ and ${\rm log}({P}^{\rm i}/{\rm day})=1.7$, respectively. Similar to the first case, the companion enters into RGB phase after about $7\times{10}^{9}\,\rm{yr}$ of evolution and begins to increase in radius. Due to the relatively short initial orbital period, a more effective tidal enhancement is exerted onto the companion, resulting in the rapid increase of mass-loss rate. The evolution of radius of the companion is shown in panel (c). As can be seen from this figure that the companion nearly fills its Roche-Lobe at the final stage (${R}/{R}_{\rm L}\sim0.905$), and the half-opening angle of the stellar wind from the companion to the WD is close to zero (${\rm sin}\alpha<0.01$), which means that the stellar wind concentrates in the equatorial disc during this stage (lasting for about $2.37\times{10}^{6}$ years).

\begin{figure}
\begin{center}
\includegraphics[width=0.75\textwidth]{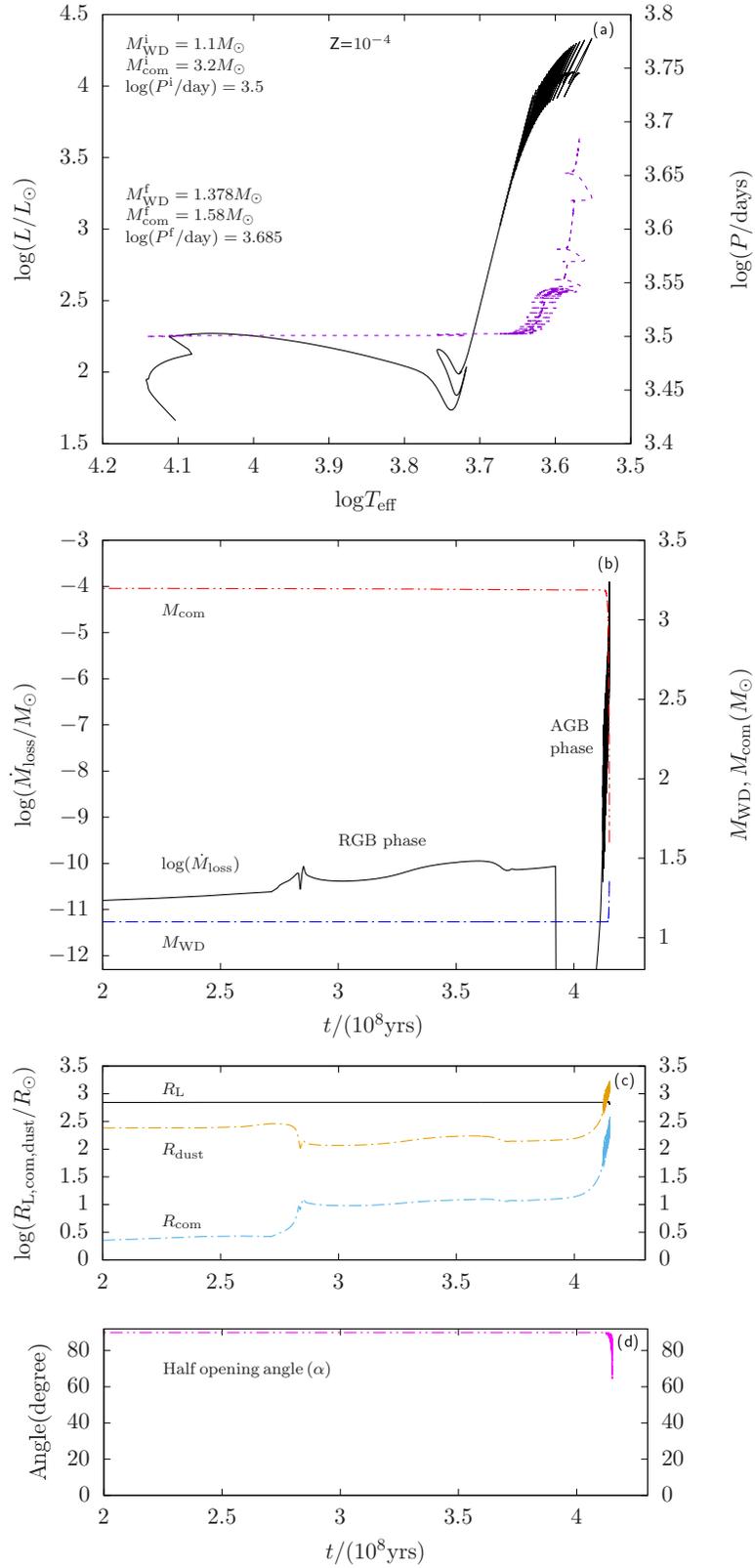}
 \caption{Similar to Fig.\,1, but for the evolution of different companion in low metallicity environment, in which the mass and initial orbital period are ${M}_{\rm com}^{\rm i}=3.2{M}_{\odot}$ and ${\rm log}{P}^{\rm i}=3.5$, respectively.}
  \end{center}
\end{figure}

A typical example of binary evolution of low metallicity population is shown in Fig.\,5, where ${M}_{\rm WD}^{\rm i}=1.1{M}_{\odot}$, ${M}_{\rm com}^{\rm i}=3.2{M}_{\odot}$, $Z=0.0001$ and ${\rm log}{P}^{\rm i}=3.5$. The companion stars usually have smaller radius and low wind-mass-loss rate compared to those with $Z=0.02$ during RGB stage due to the low opacity derived from low metallicity. However, the mass-loss rate is high enough in AGB phase since the AGB-wind is more sensitive to the luminosity. From this fiugre, we can see that the radius of companion combined with mass-loss rate increase obviously when the star enters into Thermally Pulsing AGB phase (TPAGB). In panel (c) and (d), we plotted the evolution of dust radius and half-opening angle. The dust radius ${R}_{\rm dust}$ exceeds ${R}_{\rm L}$ in the TPAGB phase, where the WD can accrete material more efficiently through the assumption of WRLOF comparing with BH. Due to the large amplitude in radial pulsation, half-opening angle of the stellar wind can vaies from about $60^{\circ}-90^{\circ}$ in this stage.

\subsection{Parameter space for producing SNe Ia}

\begin{figure}
\begin{center}
\includegraphics[width=1.0\textwidth]{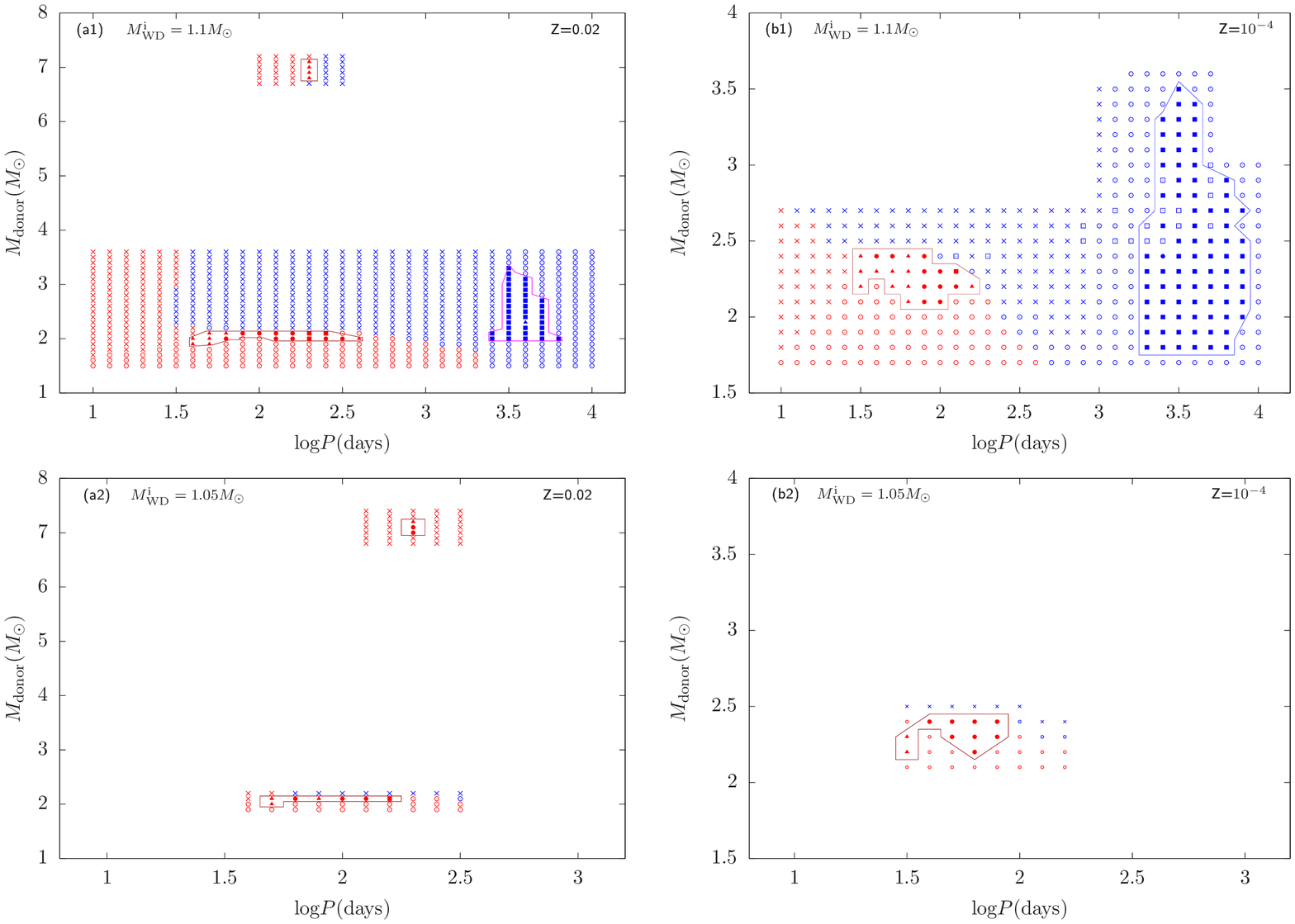}
 \caption{Regions of initial companion mass and orbital period which can produce SNe Ia with ${M}_{\rm WD}^{\rm i}=1.1{M}_{\odot}$ and $1.05{M}_{\odot}$ in solar metallicity and metal-poor environment. The open circles denote that companion mass is not enough to increase the WDs for reaching ${M}_{\rm Ch}$. Filled triangles, circles and squares represent that the WDs explode as SNe Ia in the weak H-shell flash phase, stable hydrogen burning phase and optical thick wind phase. Red and blue dots represent the core of companion stars are helium core and carbon-oxygen core when WDs explode as SNe Ia, in which the predicted ranges are enclosed by brown and magenta curves. Open squares represent the systems that come across some numerical problems when the mass donors are in TPAGB phase. The alternate burning of hydrogen and helium shells makes the calculational time-scale extremely short. For these systems, we just simply deduce that they may evolve to SNe Ia if they are in the range surrounded by blue curve, otherwise they may experience dynamical mass-transfer processes.}
  \end{center}
\end{figure}

We evolved more than 1200 WD+MS systems in solar metallicity environment, and more than 800 pairs of systems are in low metallicity environment, where the initial WD masses are from $1.0{M}_{\odot}$ to $1.1{M}_{\odot}$. The companion masses are in the range of $1.5-7.2{M}_{\odot}$ for solar metallicity and $1.7-3.6{M}_{\odot}$ for low metallicity. The initial orbital periods ${\rm log}({{P}^{\rm i}/{\rm day}})$ are from $1.0-4.0$. Fig.\,6 present the initial parameter space for producing SNe Ia in two metallicity environments under our assumption. Regions surrounded by brown or blue curves represent that the WDs can increase their masses to ${M}_{\rm Ch}$ when their companions evolve to RGB or AGB phase, respectively. Binaries beyond the left and upper boundaries would experience dynamically unstable mass-transfer and then may evolve to common envelope phase to prevent the WD from increasing its mass, whereas those beyond the right and lower boundaries tend to evolve to the double WD systems since the companions fail to provide enough masses for WDs to accrete effectively. Note that there exists possibility that binaries including a massive companion can produce SNe Ia since these companions ignite central helium non-degenerately and do not expand large enough to transfer sufficient mass before they enter into AGB phase. The parameter space shrinks dramatically with the decrease of initial mass of the CO WD, and vanished for the systems with ${M}_{\rm WD}^{\rm i}=1.0{M}_{\odot}$ CO WDs in both solar metallicity and metal-poor environment, which means that SNe Ia can only arise in the SySts that consist of massive CO WDs under current assumptions.

\begin{figure}
\begin{center}
\includegraphics[width=0.65\textwidth]{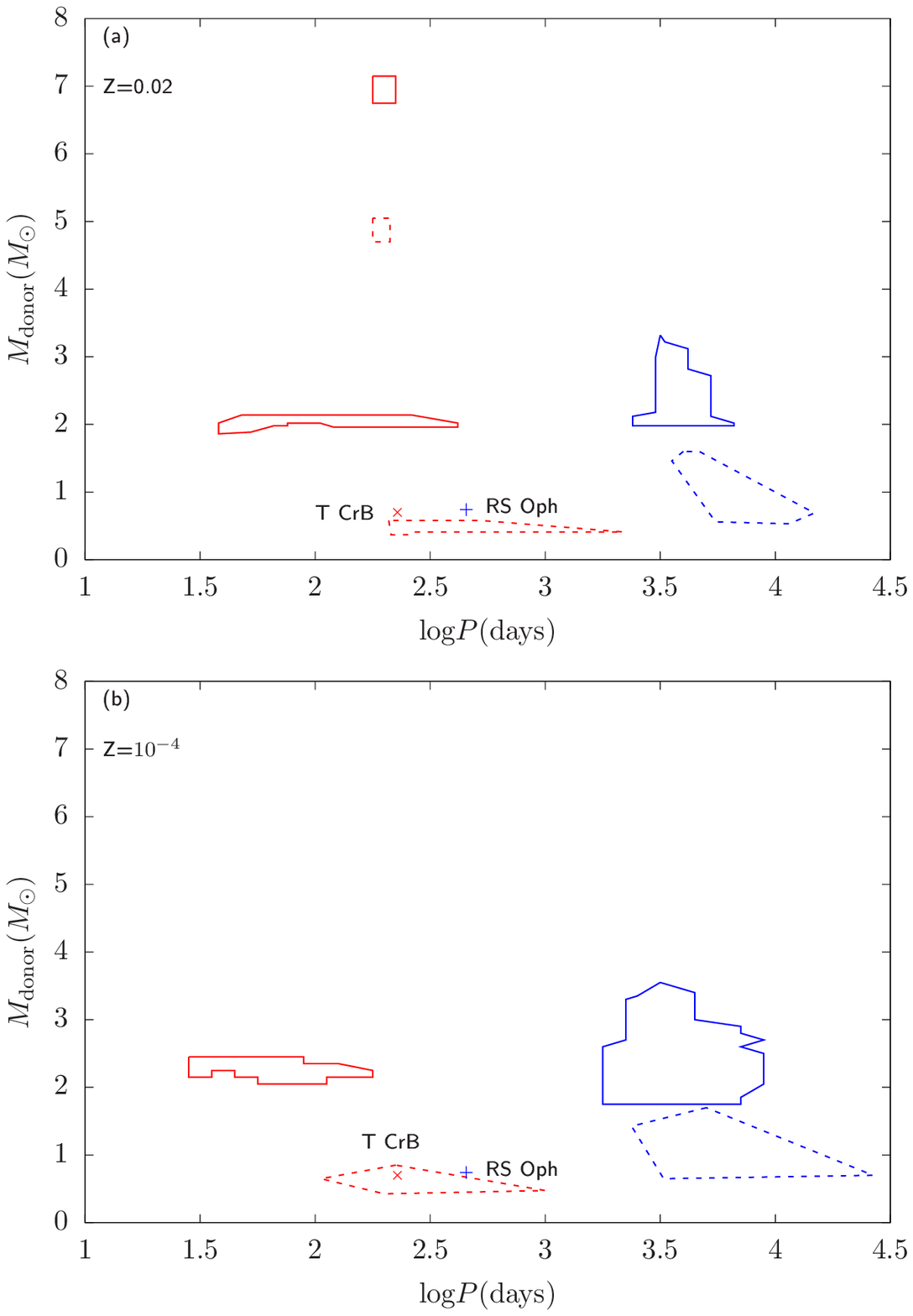}
 \caption{Regions of initial (solid curves) and final (dashed curves) companion masses and orbital periods for producing SNe Ia with ${M}_{\rm WD}^{\rm i}=1.1{M}_{\odot}$ via wind-accretion channel in solar metallicity (panel a) and metal-poor environment (panel b). Red and blue crosses represent two symbiotic systems T CrB and RS Oph, respectively.}
  \end{center}
\end{figure}

The final companion mass and orbital period distributions are shown in Fig.\,7. Comparing with the solar metallicity, the predicted parameter space for the low metallicity environment is larger, \textbf{since the opacity of the companion decreases with decreasing metallicity}, resulting in a smaller radius of companion in the giant branch. This would cause the companion more easily to avoid filling its Roche-lobe to trigger the dynamically unstable mass-transfer. Particularly, ${M}_{\rm RG}$ and ${\rm log}{P}_{\rm orb}$ of two well observed symbiotic stars, i.e. T CrB and RS Oph are located near the final range, which means that they are two promising recurrent nova symbiotic systems leading to SN Ia explosions.

\subsection{Properties of circumstellar medium}

The mass-loss rate and wind velocity are two critical parameters in estimating the CSM properties. In Figs.\,8 and 9, we present the mass-loss rates, wind velocities and half-opening angles of the stellar wind from the companions to WDs (i.e. conical streams from the giants to the WDs) in two metallicity environments at the moment of SNe explosions. Since the stellar evolution has a nuclear timescale, parameters at the time when SN explosion occurs can approximately represent the CSM properties which are close to supernova. Generally, the WD+AGB systems have larger mass-loss rate and lower wind velocity than those from the WD+RGB channel due to higher luminosity and larger radius of the AGB stars. The mass-loss rate for the companion in the WD+AGB channel is mainly concentrated in the magnitude of $10^{-5}\,{M}_{\odot}\,\rm{yr}^{-1}$, and some of the systems can have mass-loss rate greater than $10^{-4}\,{M}_{\odot}\,\rm{yr}^{-1}$. For the WD+RGB system, the mass-loss rate is in the range of $3\times{10}^{-7}-3\times{10}^{-6}\,{M}_{\odot}\,\rm{yr}^{-1}$. The half-opening angle can reflect the geometric shape of the CSM dust. The figures show that the CSM usually has larger opening-angles if the WD experiences optically thick wind phase, whereas those from stable H-shell burning or nova ejection are in a wide range. In particular, some of the symbiotic recurrent nova systems may have disc-like CSM, which are consistent with some observed systems, like RS Oph and PTF 11kx. For the supernovae exploded in the WD+AGB systems, the stellar wind typically has larger opening-angle since these systems are in wide separations. Particularly, the half-opening angles of the stellar wind for the WD+AGB systems in metal-poor environment are all larger than ${75}^{\circ}$, meaning that the CSM dust formed in low metallicity environment may be generally close to spherical symmetry geometry, which may be verified by the spectropolarimetry observations.

\begin{figure}
\begin{center}
\includegraphics[width=0.65\textwidth]{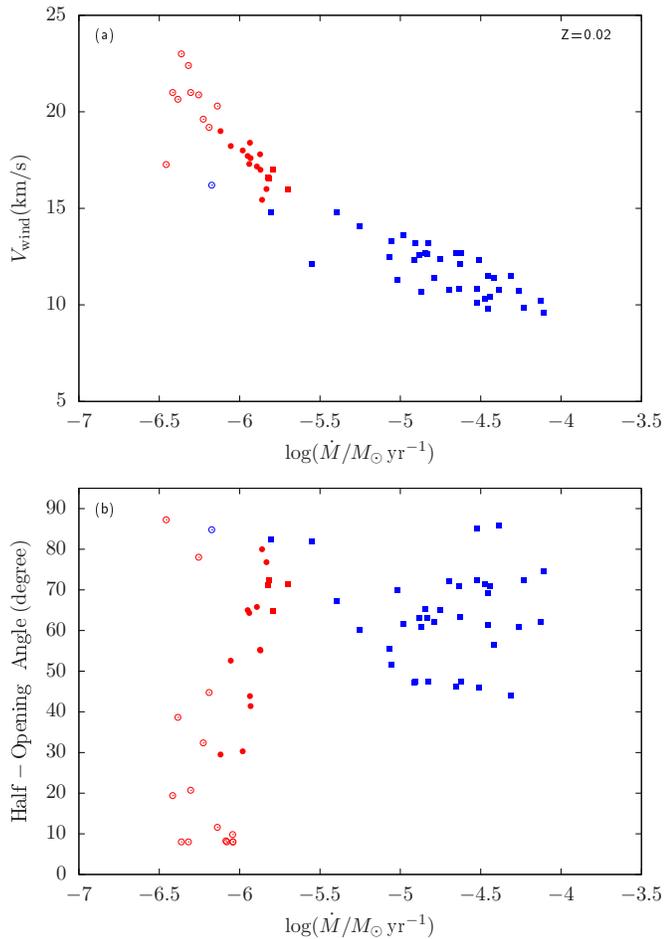}
 \caption{Panel (a): wind velocities versus mass-loss rates of the SNe Ia progenitors from the symbiotic systems in the solar metallicity at the occurrence time of supernovae explosions. Open circles, filled circles and filled squares indicate that the WDs in the corresponding systems are in the H-shell flashes, stable H burning and optically thick wind phases, respectively. Red and blue symbols represent that the supernovae explode in the WD+RGB and WD+AGB systems, respectively. Panel (b): similar to panel (a), but for the relationship between the half-opening angle of the stellar wind and mass-loss rate.}
  \end{center}
\end{figure}

\begin{figure}
\begin{center}
\includegraphics[width=0.65\textwidth]{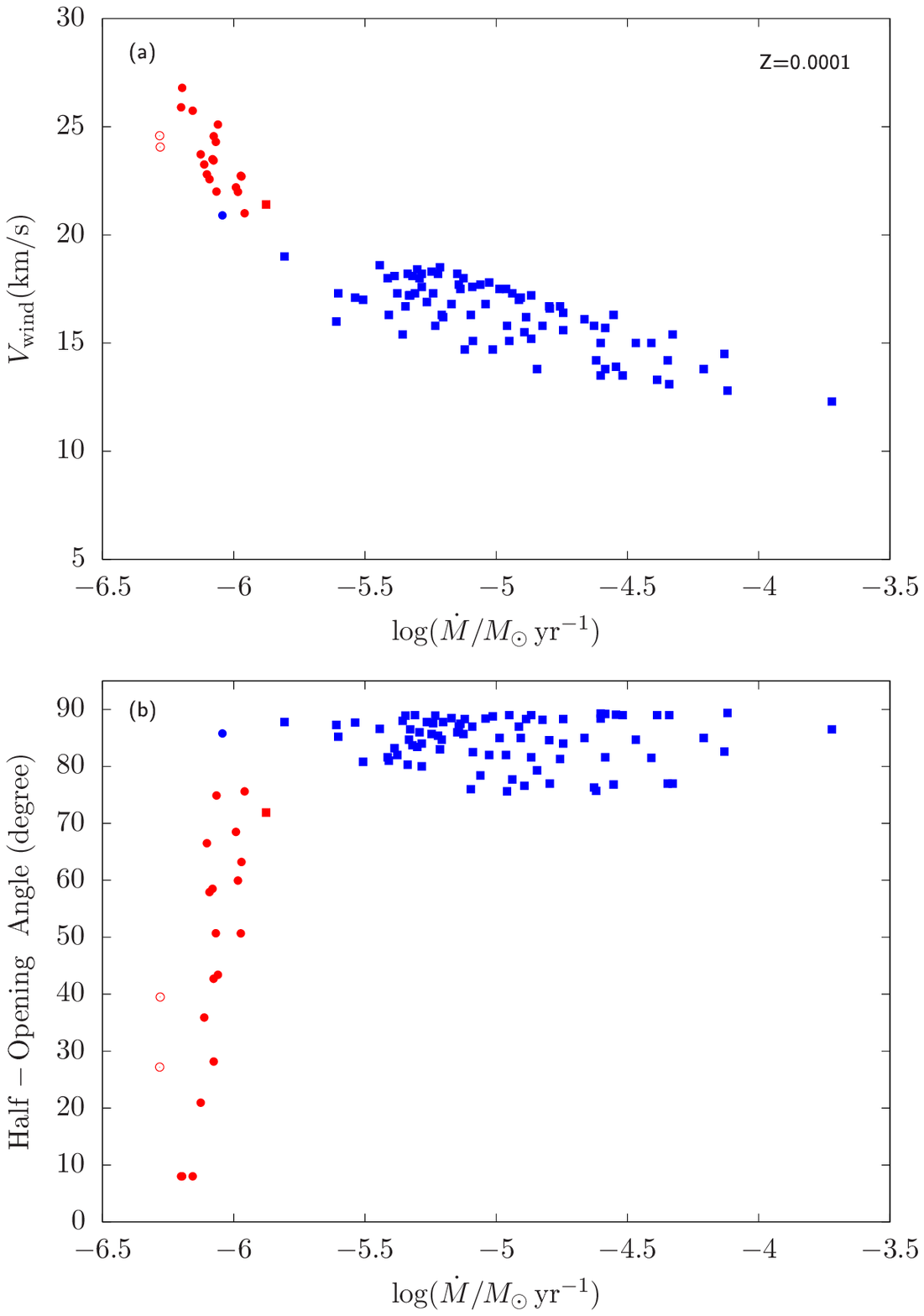}
 \caption{Similar to Fig.\,8, but for the systems in the metal-poor environment.}
  \end{center}
\end{figure}

The density profile of the CSM is a key parameter in estimating the interacting signal between the SN ejecta and CSM. In order to obtain the variation of the CSM density, we assumed a wind-like CSM, in which the stellar wind from the companion or ejected by the WD expands in a constant velocity, and the mass-loss rate is adopted from the average mass-loss rate of the progenitor in the final stage before SN Ia explosion. In this condition, the density of CSM formed via stellar wind is inversely proportional to the square of distance, i.e. ${\rho}_{\rm CSM}\propto{r}^{-2}$. For the systems with accreting WDs, the optically thick wind or nova shell ejected by the WD is able to interact with the CSM formed through the stellar wind from the companion, making the density profile of the CSM more complicated. Here we simply assumed that the interaction between WD ejection and stellar wind are completely inelastic collision, which is satisfied with the conservation of momentum, i.e. the interacted medium would have the same expanding velocity. In Figs.\,10 and 11, we estimated the possible density distributions of the CSM under our assumption, which can provide initial physics for the interaction model. The velocity of optically thick wind and nova ejecta are approximately assumed to be $1000\,\rm{km}\,\rm{s}^{-1}$ (e.g. Hachisu, Kato \& Nomoto 1999; Yaron et al. 2005; Moriya et al. 2019), and the interaction between fast wind and slow stellar wind can accelerate the CSM to a velocity of $\sim100\,\rm{km}\,\rm{s}^{-1}$, which decreases the density of CSM in the systems with high mass-loss rate. The CSM with higher density tend to appear in WD+AGB systems since the predicted wind mass-loss rate in AGB phase is greater than RGB phase; whereas for the WD+RGB channel, a dense CSM may appear in stable H-shell burning phase or weak H-shell flash phase as the stellar wind is more likely concentrated to the equatorial disk.

\begin{figure}
\begin{center}
\includegraphics[width=1.05\textwidth]{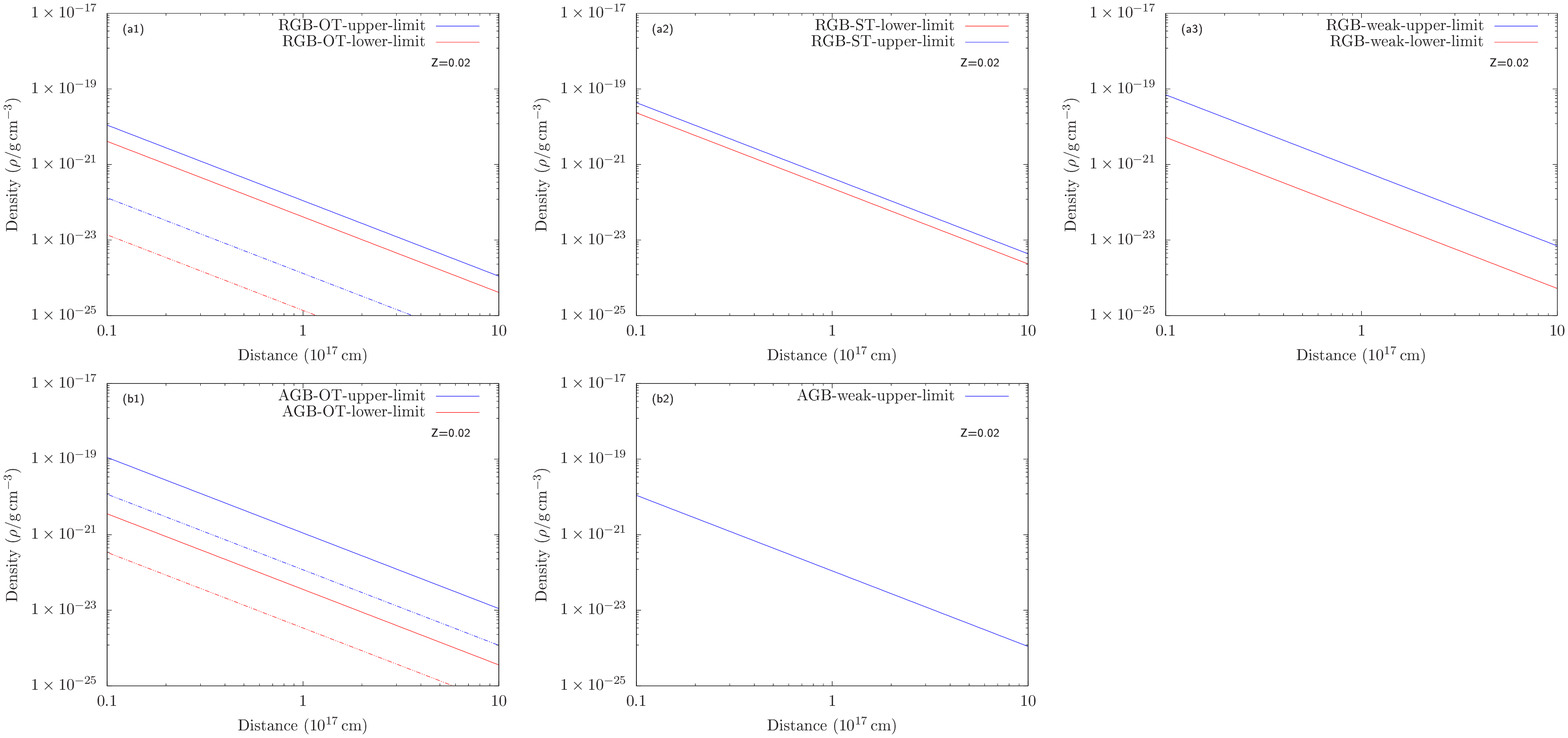}
 \caption{Density of CSM changing with distance at different phases in solar metallicity. Panel (a): WD+RGB systems in which the WDs are experiencing optically thick wind. Blue and red solid lines indicate the upper and lower limits of the density of CSM in the regions within half-opening angle of these systems, whereas the blue and red dash-dotted lines indicate the upper and lower limits in the regions within the deflection angle. Panel (b): WD+RGB systems in which the WDs are in stable H burning phase. Panel (c): WD+RGB systems in which the WDs are in the weak H-shell flash phase (here, we simply assume that the stellar wind filled up with the cavity rapidly during the interval of the recurrent nova). Panel (d): WD+AGB systems in which the WDs are in optically thick wind phase. Panel (e): WD+AGB systems in which the WDs are in the weak H-shell flash phase.}
  \end{center}
\end{figure}

\begin{figure}
\begin{center}
\includegraphics[width=1.05\textwidth]{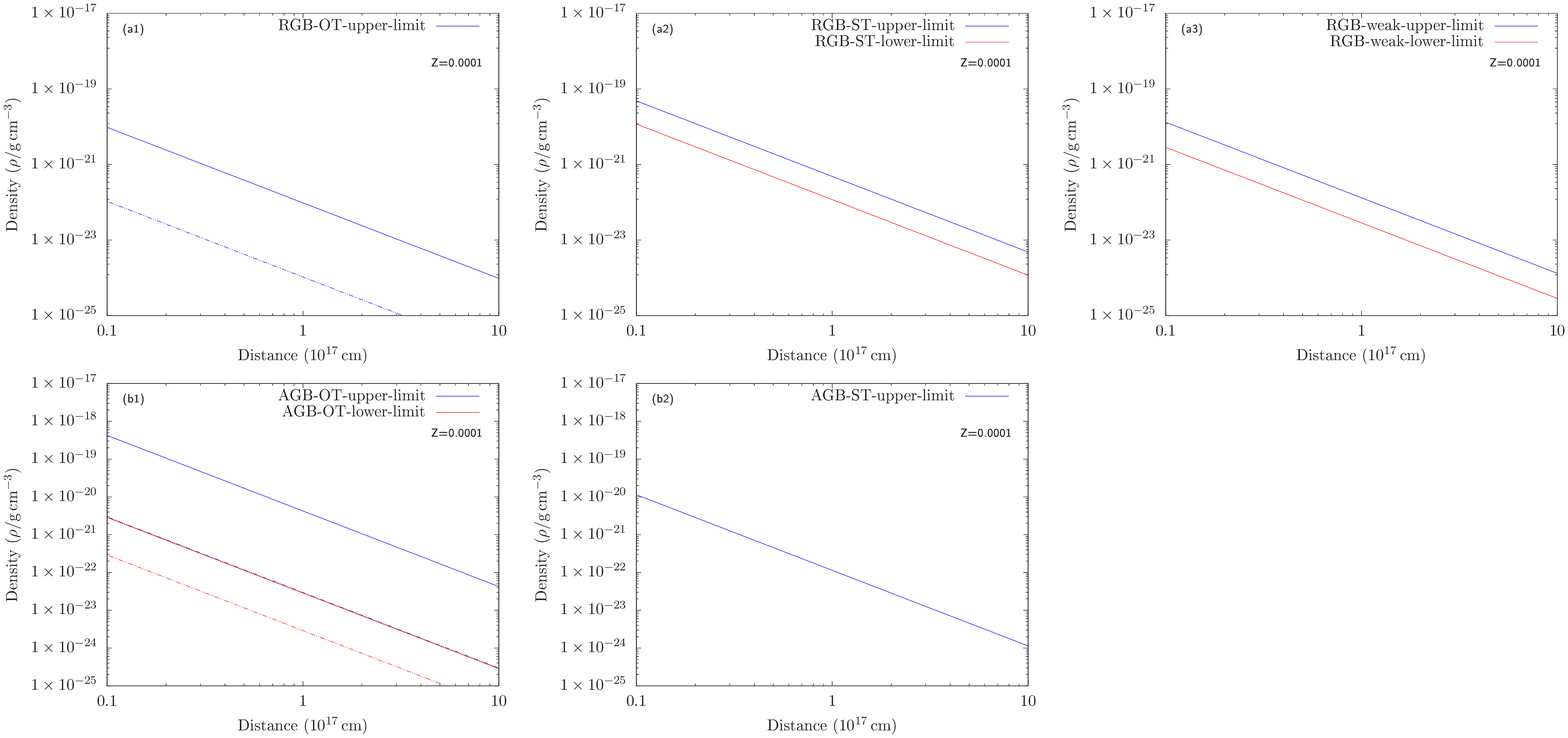}
 \caption{Similar to Fig.\,10, but for the systems in the metal-poor environment.}
  \end{center}
\end{figure}

\section{Discussion}

\subsection{Uncertainties in the modeling}

In our simulations, we assumed that the mass-loss rate of the giant donor can be enhanced via the tidal interaction. However, the degree of the enhancement is difficult to be measured, which somehow leaves large uncertainty. Chen, Han \& Tout (2011) calculated the binary evolutions under the assumption of reinforced attrition from companion (e.g. Tout \& Eggleton 1988). They found that, if assuming a significantly great value of tidally enhanced parameter (${B}\sim10000$) which estimated from Z Her, the parameter space for producing SNe Ia through wind-accretion symbiotic channel can extend to longer orbital periods and their birthrate can increase to about $6.9\times{10}^{-3}\,\rm{yr}^{-1}$. However, the enhanced effect of tidal interaction may be overestimated in the symbiotic systems. In our simulations, we assumed that both enhanced mass-loss rate and wind deflection angle are related to $({R}/{R}_{\rm L})^{6}$, which means that material lost from red giant is more concentrated towards the equatorial plane and hence increases the mass-accretion rate of the WD. Owing to the difficulties in estimating the relationship between the effect of tidal enhancement and $({R}/{R}_{\rm L})^{6}$, we assumed three different forms of ${\rm sin}\theta$, and the corresponding parameters can be estimated from the observational parameters of RS Oph as mentioned in Sect.\,2. Here, we calculate the binary evolutions of CO WD+RGB channel with linear and exponential forms of ${\rm sin}\theta$, and the parameter spaces as well as corresponding features are shown in Fig.\, 12. Comparing with panel (a1) of Fig.\,6, we can find that the parameter spaces for producing SNe Ia via WD+RGB channel (red symbols within brown curves) under the assumptions of linear and binomial functions have almost no difference, whereas the parameter space under the assumption of exponential function slightly shrinks, since the stellar wind has greater value of deflection angle in linear and binomial functions when ${R}/{R}_{\rm L}<0.7$, resulting in a higher density of stellar wind surrounding the WD. In this case, WD can accrete wind material more efficiently and thus more easily to increase in mass to ${M}_{\rm Ch}$. The distributions of SNe Ia systems in ${v}_{\rm wind}$-${\rm log}\dot{M}$ panel and ${\theta}$-${\rm log}\dot{M}$ panel are similar under different assumptions, which means that the influence of different function forms on our results are negligible.

Another uncertainty in our simulations is the limitation of mass-accretion rate of the WD. In wind-accretion process, the accretion rates of WDs are calculated by approximated equations which related to the parameters of binaries, and may overestimate the accretion rate of WDs in some cases during the evolutions. Since the WDs in symbiotic systems cannot accrete all of the wind material from their giant companions, we artificially set the upper limit of mass-accretion fractions for different accretion modes, although the proportion of angular momentum loss from RG or WD can influence the process of binary evolution and thus the parameter space for producing SNe Ia. In our calculations, we simply adopted the corresponding limitations based on previous works (e.g. I{\l}kiewicz et al. 2019; Lv et al. 2009).

\begin{figure}
\begin{center}
\includegraphics[width=1.0\textwidth]{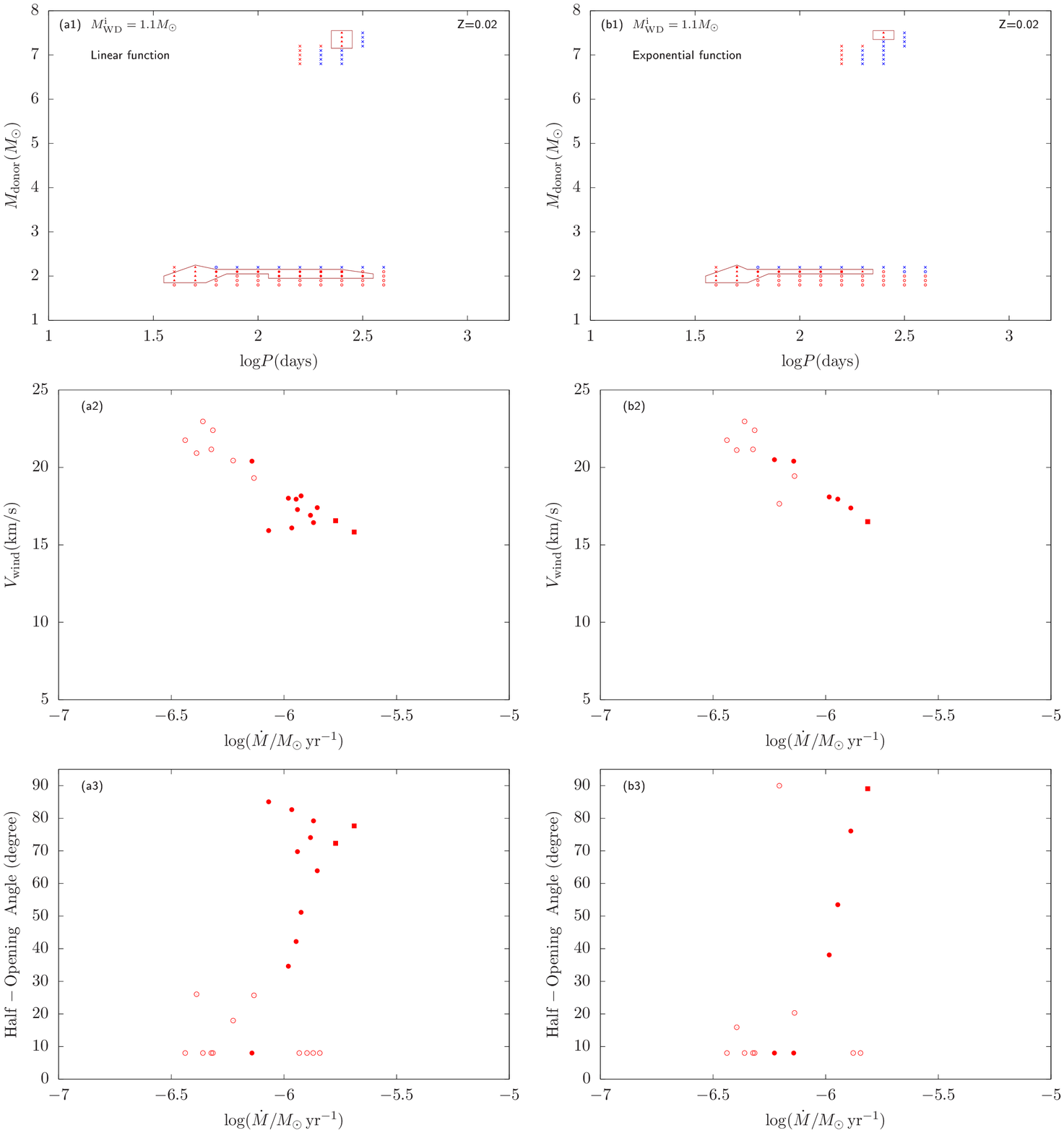}
 \caption{Parameter space for producing SNe Ia via symbiotic channel and the predicted stellar wind properties at the moment when SNe explode under two different function forms between ${\rm sin}\theta$ and $({R}/{R}_{\rm L})^{6}$. Linear function: panel (a1)-(a3); Exponential function: panel (b1)-(b3).}
  \end{center}
\end{figure}

Previous studies inferred a relatively low rate of SNe Ia from the SySts. This is because, (1) the Roche-lobe mass-transfer usually leads to irreversible expansions of the giants. The extremely fast mass-transfer from the giant donor to the WD may cause dynamically instable process, resulting in the formation of common envelope, which could prevent the WD from evolving to SN Ia in the traditional SD scenario (e.g. Li \& van den Heuvel 1997; Wang, Li \& Han 2010); (2) The wind mass-transfer usually causes an increase of the orbital separation between the WD and its giant companion, resulting in a low efficiency with which the WD retains material (e.g. Kenyon et al. 1993; Yungelson et al. 1995; Branch et al. 1995). We have not calculated the birthrate of SNe Ia via symbiotic channel, since SNe Ia can hardly stem from the binaries with initially less massive WDs under our assumptions, which predicted a relatively low birthrate. Indeed, there are some strong constraints on the symbiotic SD scenario from the observations which means that the SNe from SySt are relatively rare although they are not completely excluded. For instance, radio observations place strong constraints on the presence of CSM, and the non-detections imply a low rate of thermonuclear SNe which have symbiotic progenitors (e.g. Chomiuk et al. 2016; Lundqvist et al. 2020); The studies of supernova remnants in the Milky Way and LMC indicated the absence of luminous red giants in SN Ia progenitor systems (e.g. Li et al. 2011; Gonz\'alez Hern\'andez et al. 2012; Edwards, Pagnotta \& Schaefer 2012). On the simulations, note that some other alternative paths for producing SNe Ia via the symbiotic channel are proposed to increase the birthrate. Hachisu, Kato \& Nomoto (1999) suggested that the stellar wind from the WD may interact with the RG and strip some of the mass from the RG surface, which can stabilize the mass-transfer process and avoid the formation of common envelope. Lv et al. (2009) estimated that the Galactic birthrate of SNe Ia via symbiotic channel is between $1.03\times{10}^{-3}$ and $2.27\times{10}^{-5}\,\rm{yr}^{-1}$ by considering a spherical + equatorial disc concentrated stellar wind. By adopting an improved method to calculate the mass-transfer rate, Liu et al. (2019) found that the parameter space for producing SNe Ia via the semidetached symbiotic channel is significantly enlarged. However, none of these models can be completely proved by the observations so far, making the birthrate and circumstellar environment of symbiotic systems still be an open question. The Roche-lobe related aspherical symmetry stellar wind assumption proposed in the present work provide an additional possibility of wind accretion, in which more complicated CSM density distribution is predicted to be formed, linking the properties of ejecta-CSM interactions with the evolutionary process of the progenitors, which are hoped to be detected in the observations especially for the spectropolarimetry observations in the future (e.g. Cikota et al. 2017; Yang et al. 2018).

\subsection{Comparing with observations}

Observationally, some symbiotic novae consisting of massive WDs are thought to be the progenitors of SNe Ia. Two typical examples are RS Oph and T CrB. The symbiotic recurrent nova RS Oph consists of a $1.3-1.4{M}_{\odot}$ WD and a $0.68-0.8{M}_{\odot}$ RG companion in an circular orbit with orbital period of $\sim454.1\pm0.41\,\rm{d}$ and orbital inclination of about $50^{\circ}$ (e.g. Brandi et al. 2009; Booth, Mohamed \& podsiadlowski 2016). The WD mass in T CrB is about ${M}_{\rm WD}\sim1.37\pm0.13{M}_{\odot}$ and the mass of RG ${M}_{\rm RG}\sim1.12\pm0.23{M}_{\odot}$. The orbital period of T CrB ${P}_{\rm orb}\sim226.57\,\rm{d}$ and orbital inclination ${i}\sim67^{\circ}$ (e.g. Stanishev et al. 2004). Our calculations show that the current companion masses and orbital periods of both systems are close to the ${M}_{\rm com}^{\rm f}-{P}_{\rm orb}^{\rm f}$ range for producing SNe Ia, which indicate that RS Oph and T CrB are strong candidates in which the massive WDs in both systems can grow in mass to ${M}_{\rm Ch}$ through the wind-accretion process. Additionally, the RG in RS Oph may be close to filling its Roche-lobe by measuring its rotation period (e.g. Zamanov et al. 2007). According to our assumption, there exist symbiotic recurrent nova systems that consist of an almost Roche-lobe filling RG with disc-like stellar wind can produce SNe Ia, in which the companion mass and orbital period are close to those of RS Oph. This implies that our model is consistent with the observations.

Some SNe Ia have been reported to show apparently evidence of interaction between supernova ejecta and CSM. SN 2006X is a normal SN Ia which show clear evolution of Na I D doublet lines (e.g. Patat et al. 2007). The mean velocity of CSM is about $\sim50\,\rm{km}\,\rm{s}^{-1}$, and the circumstellar hydrogen is likely concentrated in a thin spherical shell with a radius of $\sim{10}^{17}\,\rm{cm}$. The upper limit to the shell mass can be estimated as $3\times{10}^{-4}\,{M}_{\odot}$ based on some conservative assumptions, and the dust surrounding the supernova may have multiple shells (e.g. Patat et al. 2007; Wang et al. 2008). According to our calculations, the mass-loss rates of RGs in symbiotic recurrent nova systems is around $\sim3\times10^{-7}-1\times{10}^{-6}\,{M}_{\odot}\,\rm{yr}^{-1}$ and the mass of CSM can be approximately calculated by
\begin{equation}
{M}_{\rm CSM}\approx\frac{{R}_{\rm distance}\dot{M}_{\rm loss}}{{v}_{\rm wind}}-\Delta{M}_{\rm WD},
\end{equation}
where ${R}_{\rm distance}$ is the distance between the CSM and SN Ia, and $\Delta{M}_{\rm WD}$ is the increasing mass of the WD during the time when the CSM shell expands to the corresponding distance. Assuming that the nova ejecta pushed the CSM to the distance of about ${10}^{17}\,{\rm cm}$ and formed a thin shell, we can estimate that the mass of CSM shell is about $2-8\times{10}^{-4}\,{M}_{\odot}$, which can explain the observational properties of SN 2006X. This implies that the progenitor system of SN 2006X may consist of a $\sim2.0{M}_{\odot}$ RG and a massive CO WD with an orbital period of about $40-60\,{\rm d}$, which is similar to the symbiotic system RS Oph. However, it should be mentioned that the earlier explanation of the time-variable Na I absorption lines from SN 2006X may suffer from some difficulties. Soker (2014) argued that in order to account for such variations, large hydrogen density is needed, and such multiple shells must be at a distance of larger than $0.1{\rm pc}$ from the explosion, which implied a large shell mass that cannot be explained by the SD scenario.

SN 2002ic and PTF 11kx are two SNe Ia for which clear evidence of circumstellar interaction are detected (e.g. Hamuy et al. 2003; Dilday et al. 2012). Chugai \& Yungelson (2004) estimated the mass of CSM around SN 2002ic by fitting the interacted bolometric light curve under the assumption of spherically symmetric CS envelope, and found that the total mass of CSM within $7\times{10}^{15}\,{\rm cm}$ is about $0.4{M}_{\odot}$. According to the spectropolarimetry data, Wang et al. (2004) suggested that the SN 2002ic exploded inside a dense, clumpy, disc-like circumstellar environment. PTF 11kx also show the evidence of existing massive CSM. Soker et al. (2013) estimated that the mass of CSM may greater than $0.2{M}_{\odot}$ at a distance of $\sim{10}^{16}\rm{cm}$, which is larger than what most SD scenarios predict, implying that the massive CSM may originate from common envelope ejection (i.e. PTF 11kx is stem from the merger of a WD with the core of its giant companion). Graham et al. (2017) later estimated that the total mass of CSM is about $0.06{M}_{\odot}$ within ${10}^{16}\,{\rm cm}$ by assuming a solar abundance and a spherical shell geometry for the CSM in the PTF 11kx system, depending on the reassessed of column density of Ca II. From our results, we can estimate that the CO WD+MS binary systems with initial companion mass of about $2.4-2.8{M}_{\odot}$ and initial orbital period of about $3000-6000\,{\rm d}$ can evolve into the AGB symbiotic sysytems with a mass-loss rate of AGB star in the range of $1-2\times{10}^{-4}\,{M}_{\odot}\,\rm{yr}^{-1}$, which can produce enough dense CSM comparing with the observational properties of PTF 11kx inferred from Graham et al. (2017). But for the circumstellar environment of SN 2002ic, none of the symbiotic systems can explain the dense CSM with disc-like geometry, unless the RG has a significantly high mass-loss rate (i.e. ${10}^{-3}-{10}^{-2}\,{M}_{\odot}\,\rm{yr}^{-1}$) and it almost fills its Roche-lobe simultaneously. However, binaries in this case may easily evolve to the common envelope phase through the dynamically unstable mass-transform, and may finally form double WD systems or undergo merger process which depends on whether they can survive from common envelope phase or not. Therefore, SN 2002ic may originate from some other scenarios such as the explosion of the core of an AGB star (e.g. Livio \& Riess 2003).

There exist evidences showing that metallicity has an influence on the formation of SNe Ia through symbiotic channel. Silverman et al. (2013) found that the host galaxies of SNe Ia-CSM are all late-type spiral galaxies like Milky Way with solar metallicity or dwarf irregulars like the Large Magellanic Cloud with subsolar metallicity. One exception is ASASSN-18tb, which exhibits a fairly typical light curve and has a luminosity of an underluminous or transitional SN Ia, while it is hosted by an early-type galaxy dominated by old stellar populations (e.g. Kollmeier et al. 2019; Vallely et al. 2019). Wang et al. (2013, 2019) suggested that SNe Ia with high photospheric velocities tend to have metal-rich progenitor systems and originate from symbiotic systems. Pan (2020) found that the SNe Ia with high photospheric velocities are likely formed from a distinct population which favors massive host environments, and argued that the metallicity is the only important factor in forming high-velocity SNe Ia. Our results indicated that the parameter space for producing SNe Ia through the symbiotic channel with both rich and poor metallicity dramatically shrinks with the decrease of initial WD mass. Combining with the previous results that the stars in metal-rich environment generally produce less massive WDs (e.g. Umeda et al. 1999), we can estimate that SNe Ia stem from symbiotic channel which can produce much more dense CSM gas or dust in high metallicity environment have a relatively lower birthrate than those in low metallicity environment.

Besides of the symbiotic systems, SNe Ia could also stem from the CO WD+He star systems. Moriya et al. (2019) estimated the CSM properties in CO WD+He systems, and found that this kind of system can also produce high enough density of CSM when SNe explode through ``optically thick wind'' scenario whcih can be revealed by radio observations. Li et al. (2020) investigated a sample of 16 SNe Ia, and found that all HV samples that exist in metal-rich environment have redshifted of the velocity of late-time [Fe II] features, which can be explained by the He-detonation model (e.g. Townsley et al. 2019), linking the HV SNe Ia with He star systems. Therefore, it is still early to make conclusion about the origin of CSM around SNe Ia and the effect of metallicity. Probably more SNe Ia samples of different subclasses observed in the future could provide the opportunities to reveal the corresponding questions.

\section{Summary}

By assuming that the tidal force from the WD can alter the wind mass-loss rate of the giant and the degree that the stellar wind deviate from spherical symmetry, we investigated the production of SNe Ia through the symbiotic channel. We obtained the parameter space for producing SNe Ia in both solar metallicity and metal-poor environment and analyzed the property of the binary systems including the density distribution of CSM at the moment of supernova explosions. Our results enable to reproduce some observational properties of stellar wind in SySts and CSM around SNe Ia with aspherical symmetric geometry structures, by using which one may make better constrains on the progenitors when combining with the observations of CSM interaction.

\section*{Acknowledgments}
CYW is supported by the National Natural Science Foundation of China (NSFC grants 12003013). DDL is supported by the National Natural Science Foundation of China (No. 11903075), the Western Light Youth Project of Chinese Academy of Sciences and Yunnan province (No. 202001Au070054). XFW is supported by the National Natural Science Foundation of China (NSFC grants 11325313, 11633002 and 11761141001), the National Program on Key Research and Development Project (grant no. 2016YFA0400803). BW is supported by the NSFC (Nos 11873085 and 11521303), the Chinese Academy of Sciences (No QYZDB-SSW-SYS001), and the Yunnan Province (Nos 2018FB005, 2019FJ001 and 202001AS070029).

\section*{DATA AVAILABILITY}
Results will be shared on reasonable request to corresponding author.

\label{lastpage}
\end{document}